# Role of CDW fluctuations on the spectral function in a metallic CDW system

T. Yokoya,[1,*] T. Kiss,[1] A. Chainani,[2,3] S. Shin,[1,3] and K. Yamaya[4]

[1]Institute for Solid State Physics, University of Tokyo, Kashiwa, Chiba 277-8581, Japan

[2]Institute for Plasma Research, Gandhinagar, Gujarat 382428, India

[3]The Institute of Physical and Chemical Research (RIKEN), Sayo-gun, Hyogo 679-5143, Japan

[4]Department of Applied Physics, Hokkaido University, Kita Ku, Sapporo, Hokkaido 060-8628, Japan

[*]Present address: Japan synchrotron radiation research institute (JASRI)/SPring-8, Sayo-gun, Hyogo 679-5198, Japan

**Abstract**

Angle-resolved photoemission spectroscopy of $ZrTe_3$ has been performed from room temperature($T$) down to 6 K (across $T_{CDW}$ = 63K) to study the charge-density-wave (CDW) fluctuation effects in a metallic CDW compound having Fermi surface (FS) sheets with differing dimensionality. While spectra on the 3-dimensional(D) FS show typical Fermi-Dirac function-like $T$-dependence, those along the quasi 1-D FS show formation of a pseudogap, starting at a much higher $T$ than $T_{CDW}$. Simultaneously, a van-Hove singularity consisting of the quasi 1D FS intersecting the 3D FS shows increase of coherence. This demonstrates the role of CDW fluctuations on the spectral function, and relation to the dimensionality of the states, in a metallic CDW system.

In quasi one dimensional (1D) and two-dimensional (2D) materials that exhibit charge-density-wave (CDW) order, it is known that fluctuation effects strongly influence the normal-state properties.[1] The BCS theory of superconductivity, when applied to the CDW transition, explains the opening of a CDW gap at a transition temperature($T_{CDW}$) which is equal to the mean field transition temperature($T_{MF}$). Inclusion of fluctuations reduces $T_{CDW}$ to ~($T_{MF}$/4) and predicts a pseudogap between $T_{CDW}$ and $T_{MF}$.[2,3] Angle-resolved photoemission spectroscopy (ARPES) is unique in its ability to measure the momentum($k$)- and energy($w$)- dependence of the electronic states (corresponding to the one electron spectral function A($k,w$)) of a solid. It has been used extensively to study $k$- and $w$- dependent CDW-induced electronic-structure change across $T_{CDW}$ to check the validity of the known theoretical predictions and to see the deviation from that in real materials which show a variety of physical properties.[4] However, $k$ and $w$ dependent A($k,w$) of a quasi-1D or 2D CDW material showing fluctuation effects is not known experimentally as a function of temperature($T$) upto ~$T_{MF}$.

Here, we show the role of CDW fluctuation effects on A($k,w$) in a metallic CDW system from $T$-dependent ARPES of ZrTe$_3$, which exhibits a CDW transition around 63 K, remains metallic across a bump anomaly in the electrical resistivity and shows filamentary superconductivity below 2 K.[5] Figure 1(a) illustrates the crystal structure of ZrTe$_3$, which consists of infinite rods formed by stacking ZrTe$_3$ prisms along the $b$ direction.[6] The electrical resistivity in the metallic phase is anisotropic with ($r_a$ : $r_b$ : $r_c$ = 1: 1: 10) and surprisingly, only $r_a$ and $r_c$ exhibit the resistivity anomaly.[5,7] This is in contrast to isostructural NbSe$_3$, which is also a quasi-1D CDW compound but with a resistivity anomaly along the prismatic rods *i.e.* the $b$ direction in Fig. 1(a).[8] The resistivity anomaly in ZrTe$_3$ is accompanied with Hall coefficient[5] and thermopower changes[7] but a direct measurement of superlattice spots with a vector $q = 0.93a*+0.33c*$ from an electron microscopy study revealed the CDW vector in the *a-c* plane.[9] From an analysis of the structural data, which indicate ditelluride groups, a naïve valence description of the type $(Zr^{4+})(Te^{2-})(Te_2^{2-})$ and $(Zr^{2+})(Te^{2-})(Te_2)$ is obtained from electron counting.[7] It is understood that the real picture is intermediate valency between $Zr^{4+}$ and $Zr^{2+}$, with a reduction-oxidation (redox) mechanism in operation which emphasizes covalency. Two independent band-structure calculations of ZrTe$_3$ have predicted a 3D FS sheet around the G point, quasi-1D FS sheets along the B(A)-D(E) direction, and a vHs on B-A line where hybridization of Zr 4*d* and Te 5*p* derived states add a more 3D character to the quasi-1D sheet (thus "dual quasi-1D+3D").[7,10] The calculated FS sheets intersecting the Brillouin zone (BZ) in the $k_z = 0$ plane are shown as thick blue lines in Fig. 1(b). The calculated quasi-1D FS sheet forms a nesting vector consistent with the CDW vector obtained by the electron microscopy study[9]. The material is thus well-suited to study fluctuation effects in an accessible *T* range from below $T_{CDW}$ to above $T_{MF}$ and compare it with dimensionality (quasi-1D versus 3D) of the electronic states

    Single crystals of ZrTe$_3$ were prepared by the iodine transport method[5]. ARPES measurements were performed on a spectrometer built using a Scienta SES2002 electron analyzer and a GAMMADATA high-flux discharging lamp with a toroidal grating monochromator. The energy and angular resolution using He Iα (21.218 eV) resonance lines were set to ~ 15 meV and ± 0.1 deg (corresponding to 0.0067 Å$^{-1}$), respectively, to obtain reasonable count rates. Samples were cooled using a flowing liquid He refrigerator with improved thermal shielding. Sample *T* was measured using

a silicon-diode sensor mounted below the samples. The base pressure of the spectrometer was better than 5 x $10^{-11}$ Torr. The sample orientation was measured *ex-situ* using Laue photography and checked *in-situ* by symmetry of ARPES spectra. All measurements have been done for fresh surfaces obtained by *in-situ* cleavings and $T$-dependent spectral changes were confirmed by cycling $T$ across the range of reported measurements. The normalization of the $T$-dependent spectra at any measured $k$ point was performed with scan time. The Fermi level ($E_F$) of samples was referenced to that of a gold film evaporated onto the sample substrate with an accuracy of better than ± 1 meV.

We first discuss the experimental band structure near $E_F$ and FS of $ZrTe_3$ with five ARPES intensity plots (Figs.2 (a1)-(a5)) measured at 300 K. Along the G(Z)-Y(C) direction (Fig.2 (a1)), we observe two bands dispersing symmetrically with respect to the G(Z) point: the high intensity peak shows parabolic dispersion with a minimum binding energy at 0.25 eV and the low intensity peak shows a linear dispersion with $E_F$ crossings. The latter band forms a hole character FS sheet around the G(Z) point, as is evident from the intensity map at $E_F$ over the BZ corresponding to FS (Fig. 1(b)). In Fig. 2 (a3), the map has a higher intensity region near the B(A) point and just below $E_F$. This suggests that the bottom of a band is located at the B(A) point near $E_F$. The same band shows opposite curvature of dispersion along the zone boundary (Fig.2 (a2)), thus forming a van-Hove-singularity (vHs) very close to $E_F$ at the B(A) point. The bottom of the band forming the vHs remains within 0.3 eV of $E_F$ along B(A) to D(E) (Figs. 2 (a3)-(a5)), and the band makes electron-like quasi-1D electronic states (Fig. 1 (b)). The experimentally obtained band structure and expected FS topology agree with band structure calculations which predict that the FS of $ZrTe_3$ consists of a 3D hole-character

FS around the G point and quasi-1D electron-like FS sheets along the zone boundary (Fig. 1(b)), though there are small discrepancies in the energy position of bands near $E_F$. The quasi-1D electron-like FS sheets have a dominant Te $5p_x$ character originating in the Te(2)-Te(3) chain (Fig. 1(a)) and has been speculated to be responsible for the CDW transition.[7,10] The coexistence of both hole- and electron-like sheets with relatively small area indicates the semi-metallic nature of $ZrTe_3$, consistent with optical conductivity measurements[11].

The results described above make $ZrTe_3$ extremely important as a material allowing a simultaneous study of coexisting quasi-1D and 3D FS's, along with a region in *k*-space exhibiting vHs where the FS's of differing dimensionality overlap and/or mixed. Utilizing the *k*-resolving capability of ARPES, we study the *k*- and *T*-dependence of A(*k*, *w*). The five ARPES intensity plots measured at 6K are shown in Figs.2 (b1)-(b5). Comparing 6 K plots((b1)-(b5)) with those at 300K((a1)-(a5)), one can observe that plots (a1) and (b1), corresponding to the 3D FS derived bands do not show *T*-dependent spectral changes. In contrast, plots (b2)-(b5) measured at 6K show clear spectral changes when compared with 300K plots (a2)-(a5). The most important change, which is a transfer of spectral intensity from the region near $E_F$ to higher binding energy, is seen in the plots of (b4) and (b5) when compared with (a4) and (a5). The data indicate significant electronic structure changes across $T_{CDW}$ taking place on the 1D derived states, and negligible changes at the 3D derived states.

A comparison of the *T*-dependent changes in energy distribution curves (EDCs) at and near FS crossings with three *k* points is shown in Fig.3.[12] The 3D-FS EDCs (top) is consistent with the Fermi-Dirac (FD) function change. Dramatic differences between 6 K and 300 K are observed only in the electron-like quasi-1D FS sheets for the vHs at B(A) point (middle) and the quasi-1D FS at D(E) point (bottom). These changes are related to the CDW vector connecting partially nested FS's (Fig.1b, black arrows), and which do not involve $b^*$. This is consistent with the fact that the resistivity anomaly is not observed along $b$ direction in real space. The CDW pseudogap of the electron-like FS sheets indicates a decrease of electron number, consistent with the increase of negative Hall coefficient below $T_{CDW}$ (Ref. 5). We would like to emphasize that the $T$-dependent changes are intrinsic as the normalization performed in the present study also results in spectral weight conservation within a binding energy region of ~1eV, for each $k$ point. Thus, the electronic-structure changes show clear dependence between the two types of FS sheets as well as $k$-dependence within the quasi-1D sheet: a spectral weight-transfer at D(E) point and increase of intensity at $E_F$ at B(A) point indicative of formation and absence of a gap, respectively. The negligible changes on the 3-D hole-like FS sheet is naturally explained from the expectation that the CDW-induced changes occur only on FS sheets connected with the nesting vector. The $k$-dependent spectral changes of the quasi 1D states agree well with the band calculation results of Stowe et al.[10] which predict better parallelism of the quasi-1D FS sheets for $\pi/b < k_y < 2\pi/b$ in the first Brillouin Zone. The better parallelism induces CDW nesting instabilities and thus promotes the CDW gap opening at and near D(E) ($k_y = 2\pi/b$), *i.e.* over only part of the FS (Fig.1b, black arrows). In contrast, near the B(A) point, hybridization of Zr 4$d$ and Te 5$p$ derived states add a more 3D character to the same quasi-1D sheet, giving rise to a deviation from an ideal nesting condition. The present results are reminiscent of a $k$-dependent CDW gap due to imperfect nesting as was recently reported for quasi-2D CDW compound $SmTe_3$ at room temperature(which has an estimated $T_{MF}$~1300 K but exhibits no resistivity anomalies between 1.2-300 K)[13] and quasi-1D compound NbSe$_3$ at 15 K (with an estimated $T_{MF}$ ~ 580 K and two CDW transitions at 59 K and 145 K)[14].

Considering that the top, middle and bottom panels of Fig. 3 represent T-dependent A(k,w) across $T_{CDW}$ at k-points with differing dimensionality: (i) 3D, (ii) dual quasi-1D +3D, and (iii) quasi-1D, respectively, it is interesting that the top and middle panels have changes over the same energy scales due to the 3D character. It is tempting to attribute the rather large spectral intensity changes in the middle panel to the quasi-1D character as in the bottom one, although the point B(A) has dual character.

The agreement of the experimental observations with the band calculation predicting a CDW transition induced by FS nesting in ZrTe$_3$ motivated us to measure detailed T-dependent studies over a wide T range from below $T_{CDW}$ to above $T_{MF}$. Figure 4 (a) bottom panel shows the T-dependent EDC's between 6 K and 280 K of the quasi-1D FS at the D(E) point in k-space. On decreasing T from 280 K, the spectra show decreasing spectral weight at and within 70 meV of $E_F$ with a broad peak forming at about 0.15 eV. The top panel shows T-dependent EDC's divided with corresponding FD functions to show spectral function changes without FD function.[15] We find that a clear pseudogapping starts below 200 K, a T much higher than $T_{CDW}$ and continues systematically across $T_{CDW}$, thus confirming the role of CDW fluctuations on A(k,w). The observation that the pseudogap fills-in with the peak position relatively T-independent suggests that the amplitude of the CDW in ZrTe$_3$ does not change significantly as a function of T, similar to the tunnelling spectra for quasi 1D NbSe$_3$ that also does not follow a BCS-like gap behaviour in the metallic CDW phase.[16] ARPES studies on NbSe$_3$ show evidence for fluctuation effects at 300K and suppression of spectral weight near $E_F$,[14,17] although T-dependence of the CDW fluctuation effects was not addressed. Moreover, in ZrTe$_3$, the spectra always show a residual Fermi edge indicating that the system is not like a fully gapped Peierls' system. Therefore, the observed spectral changes represent $T$-dependent A($k,w$) induced by purely CDW fluctuation effects and are different from the results on quasi 1-D compounds[15-19] including electron-electron correlation effects. However, even in the present case, the spectral weight transfer is over nearly 250 meV, which translates into ~10x$T_{MF}$, a relatively large energy scale compared to the expected scale of ~3x$T_{MF}$ from theory[2,3].

Simultaneously, the EDC's obtained at the vHs point in $k$-space at B(A) do not show gapping, but show a systematic increase in the spectral weight within 100 meV of $E_F$ and a narrowing of the peak width, indicative of increasing coherence in the quasiparticle peak. One possible origin for increasing coherence may be the electron phonon coupling, as was shown for the case of a Mo(110) surface state.[22] The strong electron-phonon coupling induces a kink in the dispersion and results in a two-peak structure in EDCs, as well as a decrease in the width of EDCs at $E_F$. In the present case, though the decrease of the width in the EDCs is observed, we do not observe a kink feature within an energy corresponding to the Debye temperature of ZrTe$_3$ [150K (Ref. 23)], nor the characteristic two-peak structure. A similar signature of a 2D FS exhibiting quasiparticle coherence has been recently shown to be also related to dimensional crossover from 2D to 3D in layered materials.[24,25] In the present case, we note that the spectral changes are very different as a function of $T$ for the quasi-1D case with pseudogapping and increasing coherence at the vHs with dual quasi-1D and 3D character. The energy scale of the spectral changes is also significantly different with the quasi-1D case exhibiting changes over larger energy scales. Fig. 4 (c) shows a plot of the spectral intensity changes as a function of $T$ at $E_F$ being consistent with fluctuation effects and the CDW transition at the point D(E). The simultaneous observation of increasing coherence at point B(A) beginning at a much higher $T$ than $T_{CDW}$ and continuing below $T_{CDW}$ is rather interesting with no precedence. Fig. 4 (d) shows the peak full-width at half maximum (FWHM) dependence on $T$ obtained by curve-fitting the quasiparticle peak to a Lorentzian function with a Shirley background and multiplied with a resolution broadened FD function. We find that the peak width reduces as $T$ is decreased, with a similar $T$-dependence as for intensity at $E_F$ for the D(E)-point, suggesting a coupling with the pseudogap formation. It is also more pronounced below $T_{CDW}$. Increase of coherence below $T_{CDW}$ has been reported from ARPES studies of the 2D CDW system 2$H$-TaSe$_2$,[26] and its relation to the CDW gap opening[27] was also discussed. Optical studies have also shown an increase of conductivity at low energies.[28] In the high-$T_c$ cuprates, increase of coherence along the nodal direction[29] and formation of a pseudogap around ($\pi,\pi$) in the BZ[30] are well-known. But the origin of the pseudogap in the cuprate is not established and its relation to fluctuating charge and/or spin order is under active discussion.[31,32]

In conclusion, we reported detailed $T$ dependent ARPES studies of ZrTe$_3$, which has quasi 1D and 3D FS. The results show simultaneous formation of a pseudogap and increase of quasiparticle coherence below $T_{MF}$, representing the role of CDW fluctuation effects in a metallic CDW system.

**Acknowledgements:** We wish to thank the late Prof. T. Sambongi for supplying single crystals of ZrTe$_3$. We thank S. Watanabe and K. Yonemitsu for very valuable discussions. This work was supported by grants from the Ministry of Education, Culture and Science of Japan.

[12] It is noted that 3D states will have a large final state broadening, resulting in a smearing of spectral function. However, we discuss the $T$-dependent spectral changes with respect to the $k$ points, particularly because of the clear difference in energy scales and spectral weight changes which depend on the $k$ points.

**Figure captions**

Fig. 1 (Color online). (a) A schematic crystal structure of ZrTe$_3$. Prismatic rods run along the *b* axis. (b) An ARPES intensity plot corresponding to experimental FS sheets, intersection of calculated FS sheets and BZ of ZrTe$_3$ with $k_z = 0$ (Ref. 10). Black arrows correspond to $a^*$ component of the CDW vector.

Fig. 2 (Color online). (a1-a5) and (b1-b5), ARPES intensity plots along gray lines (l1-l5) in Fig.1(b) measured at 300K(normal phase) and 6K (CDW phase), respectively, normalized with scan times. White lines for 300K data are calculated band dispersions. The intensity increases from black-to-red-to-yellow.

Fig. 3. *k*-dependent ARPES spectra measured at 300 K and 6 K for three FS crossing points indicated in the BZ (inset) corresponding to differing dimensionality : (top) 3D FS, (middle) dual quasi-1D + 3D, and (bottom) quasi-1D.

Fig. 4 (Color online). *T*-dependent (6 - 300 K) ARPES spectra normalized for intensity with scan time are shown for (a) the quasi-1D FS showing a fluctuation induced pseudogap at the D(E) point in *k*-space and (b) quasiparticle coherence at the B(A) point which corresponds to the vHs with dual quasi 1D and 3D character. The top panel of (a) is FD function subtracted spectra of the bottom panel of (a). (c) *T*-dependent intensities at $E_F$ due to (i) the pseudogap (red circles) at the D(E) point, and (ii) increasing quasiparticle coherence (blue diamonds) at the B(A) point. (d) The quasiparticle peak FWHM showing the reduction in peak width, clearly below 200 K and across $T_{CDW}$.

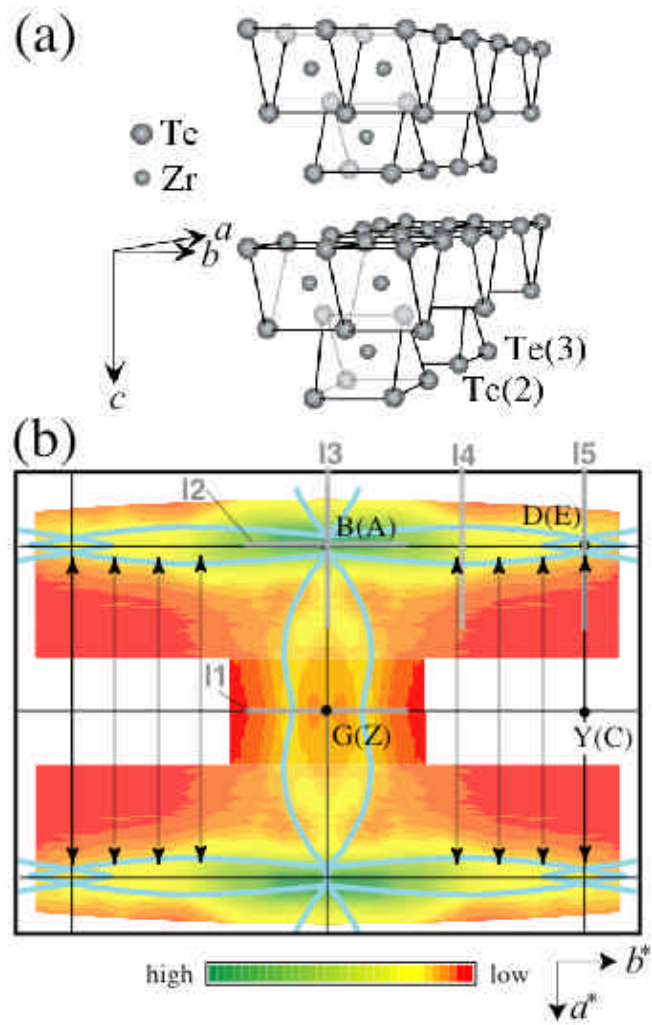

Fig.1, T. Yokoya et al.

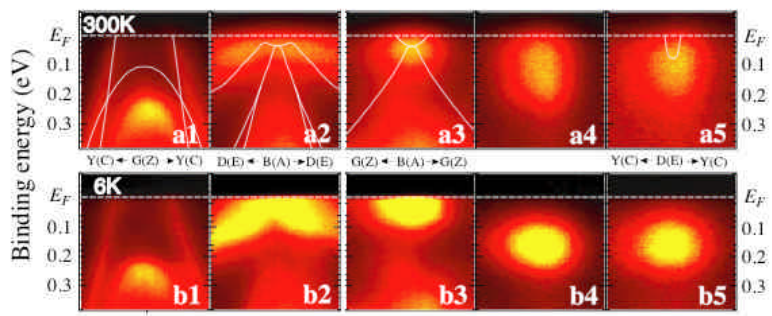

Fig.2, T. Yokoya et al.

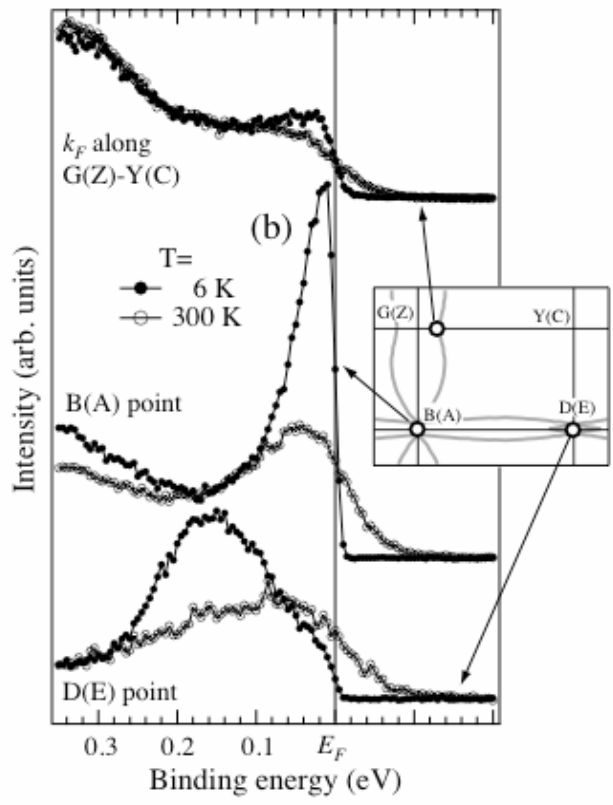

Fig. 3, T. Yokoya et al.

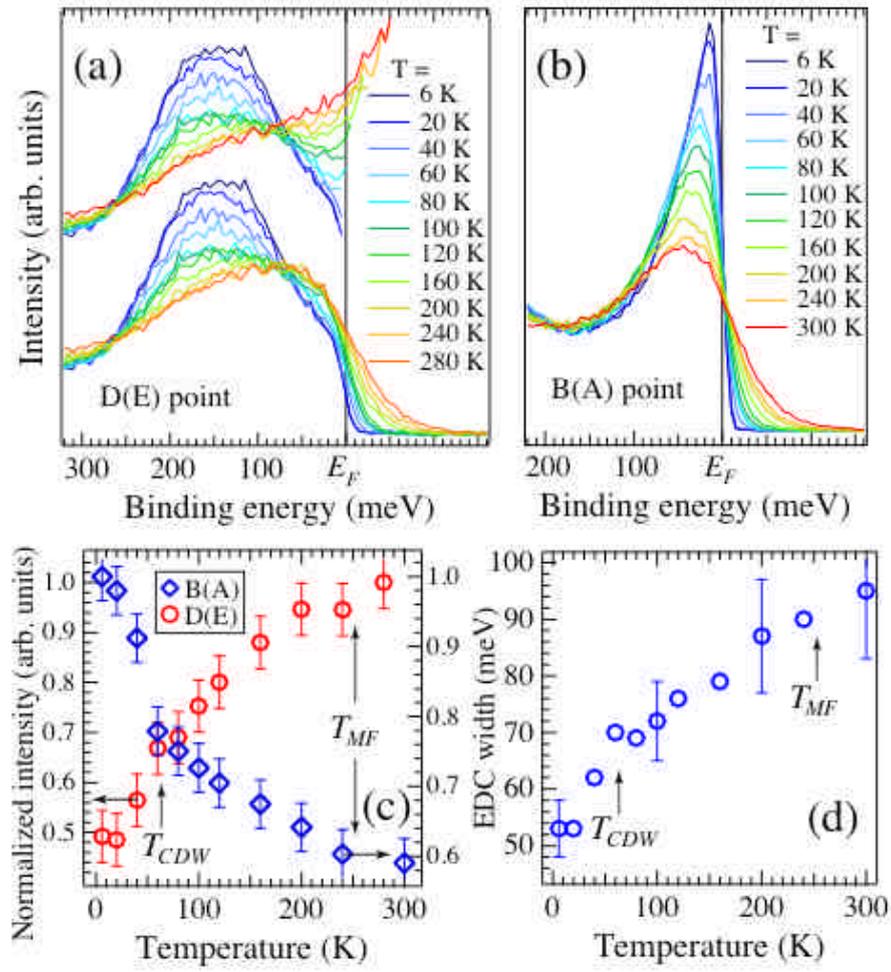

Fig. 4, T. Yokoya et al.